\documentclass{article}

\newcommand{\sys}{SWaRL}
\usepackage[ruled,vlined]{algorithm2e} 
\usepackage{amsmath}        
\usepackage{graphicx}
\usepackage{enumitem}
\usepackage{wrapfig}
\usepackage{subcaption} 
 \usepackage[preprint]{neurips_2026}

\usepackage[utf8]{inputenc} 
\usepackage[T1]{fontenc}    
\usepackage{hyperref}       
\usepackage{url}            
\usepackage{booktabs}       
\usepackage{amsfonts}       
\usepackage{nicefrac}       
\usepackage{microtype}      
\usepackage{xcolor}         

\title{SWaRL: Safeguard Code Watermarking via Reinforcement Learning}

\author{
Neusha Javidnia \thanks{Corresponding author: \href{mailto:njavidnia@ucsd.edu}{njavidnia@ucsd.edu}} \\
ECE Department\\
UC San Diego
\And
Ruisi Zhang \\
ECE Department\\
UC San Diego
\And
Ashish Kundu \\
Cisco Research
\And
Farinaz Koushanfar \\
ECE Department\\
UC San Diego
}

\begin{document}

\maketitle
\begin{abstract}
We present \sys{}, a robust and fidelity-preserving watermarking framework designed to protect the intellectual property of code LLMs by embedding unique and verifiable signatures in the generated program. Existing watermarking approaches either rely on handcrafted code transformations or manipulate token generation probabilities at inference time, making them vulnerable to removal attacks or prone to breaking functional correctness. To address these challenges, \sys{} employs a reinforcement learning-based co-training framework that uses compiler feedback for functional correctness and a jointly trained confidential verifier as a reward signal to maintain watermark detectability. Furthermore, \sys{} employs low-rank adaptation (LoRA) during fine-tuning, enabling efficient integration of watermarking behavior and transferability across model updates. Extensive experiments show that \sys{} achieves strong watermark detection accuracy compared to prior methods while fully maintaining watermarked code functionality. Moreover, \sys{} exhibits strong resilience against refactoring and adversarial transformation attacks, which maintains reliable attribution without substantial computational overhead.
\end{abstract}

\section{Introduction}

Large language models are capable of generating high-quality source code, which has tremendously accelerated productivity in software development. Building these powerful tools requires substantial investments in data curation, post-training alignment, and architecture design, all of which constitute valuable intellectual property (IP). Such high-quality code, however, can be easily redistributed, modified, or reused without attribution, obscuring the provenance of the model and undermining the ownership rights of model developers. At the same time, the ease of producing functional code at scale also raises serious concerns regarding code accountability. Malicious adversaries can exploit code LLMs to create obfuscated or vulnerable code snippets, making it difficult to trace their origins during forensic analysis. As a result, watermarking, the process of embedding imperceptible yet verifiable signatures into generated code, emerges as a crucial mechanism for tracing authorship, ensuring IP protection, and enabling accountability in the era of AI-assisted programming.

Prior code watermarking work can be generally categorized into two approaches: (i) inference-based watermarking, and (ii) neural-based watermarking.  Inference-based watermarking encodes watermarks at the token generation step by constraining the decoding process. The vocabulary is split into ``green'' and ``red'' token sets, and biasing the model to sample primarily from the green set \cite{lee-etal-2024-wrote} \cite{pmlr-v202-kirchenbauer23a}. While such methods require no additional model updates, they often degrade code fidelity as small probability bias can collapse the code syntax, thereby compromising its functionality. Neural-based watermarking, in contrast, embeds the watermark globally by training an additional watermarking network that learns to transform the code to watermarked variants with hidden signatures embedded \cite{zhang2025robustsecurecodewatermarking} \cite{10646683}. Such transformations are done with manually defined syntactic structure rules and learned variable renaming patterns. This approach can better preserve the code's functionality, but relies on manually defined transformation rules and is therefore vulnerable to refactoring attacks that systematically restructure the artifact.

We presents \sys, a robust watermarking framework that aligns code LLMs to automatically generate fidelity-preserving code with the owner’s signature embedded. \sys{} fine-tunes the model so that satisfies two objectives: (i) produce functional code that successfully passes test cases, and (ii) embed signatures that can be reliably detected by a watermark detector. Both criteria can be verified automatically without human-in-the-loop inspection or manually-crafted transformation rules, which ensures \sys's generalizability toward new syntactic structures and robustness toward attacks.

To accomplish this, \sys{} fine-tunes the model using Group Relative Policy Optimization (GRPO) \cite{shao2024deepseekmath}, a lightweight and reinforcement learning based alignment algorithm. Given an input prompt, the model samples multiple candidate code outputs, each of which is evaluated using functional unit tests for correctness and a trainable verifier for watermark detectability. To ensure stable optimization across these objectives, \sys{} employs a co-training scheme in which the watermark detector, which serves as the reward model, is jointly optimized with the actor. This coordinated update schedule prevents any single reward component from dominating and enables the model to simultaneously improve code quality and watermark strength without collapsing toward a single objective.
To keep the process lightweight and maintainable across model versions, \sys{} uses Low-Rank Adaptation (LoRA) \cite{hu2021lora} to inject watermark-specific behavior as a small adapter on top of the base LLM, rather than fine-tuning all model parameters, which ensures computation efficiency. As a result, \sys{} produces semantically correct code that carries a verifiable watermark, with modest training overhead.  

In brief, our contributions are summarized as follows:

\begin{itemize}[leftmargin=*]
    \item We introduce \sys, a reinforcement learning based watermarking framework that automatically aligns code LLMs to generate functionality correct code while embedding watermark signals, without requiring manually crafted transformation rules.
    \item \sys{} leverages (i) Group Relative Policy Optimization to align the model to generate code with functional correctness, reliable watermark detectability, and robustness against refactoring-based attacks, and (ii) Low-Rank Adapters to provide a modular and lightweight mechanism for integrating watermarking behavior into base LLMs.
    \item \sys{} delivers state-of-the-art watermark detectability while achieving the highest code-generation accuracy among existing watermarking baselines. Across all four benchmarks, \sys{} consistently surpasses the non-watermarked supervised fine-tuned (SFT) baseline in pass rate and outperforms prior watermarking methods. Furthurmore, It demonstrates the strongest robustness to refactoring attacks, exhibiting  6.4\% average AUROC degradation across evaluated datasets.
\end{itemize}

\section{Background and related work}
\subsection{Code LLMs watermarking}

Code large language models (LLMs)~\cite{touvron2023llama,chen2021codex,wang2021codet5} are increasingly utilized in the programming language domain to assist software development, e.g., code generation~\cite{copilot,luo2023wizardcoder,pan2023understanding,roziere2021leveraging}, code summarization~\cite{ahmed2022few,gao2023makes}, and automated testing~\cite{yu2023llm,kang2023large}. These models aim to learn a probabilistic mapping \(p(c \mid x)\) that predicts a target code snippet \(c\) conditioned on a context \(x\), which may include natural-language descriptions, partial code fragments, or other program artifacts. Pretraining is performed using a next-token prediction loss over large code
corpora, maximizing the likelihood of each sequence under the model. Given a dataset \(D = \{c_1, c_2, \ldots, c_N\}\), the standard training objective is the negative log-likelihood:

\begin{equation}
\mathcal{L} = -\frac{1}{N}\sum_{i=1}^{N}\log p(c_i)
\end{equation}

Cloud-based code generation APIs such as OpenAI CodeX~\cite{chen2021evaluating}, GitHub Copilot~\cite{copilot}, and Code LLaMA~\cite{touvron2023llama} now produce high-quality program synthesis at scale. As these services become widely integrated into software development workflows, effective watermarking is essential to protect provider's intellectual property and to detect unauthorized or harmful use of LLM-generated code, including plagiarism and code-based vulnerability exploitation.

Adding watermarks in LLM-generated code can be categorized into~\cite{tang2023science}: (i) rule-based watermarking~\cite{li2023protecting} uses a syntax-replacement table to insert watermarks by systematically replacing certain code tokens or constructs with the watermarked variants. It relies on handcrafted rules and language-specific heuristics to ensure successful watermark insertion.
(ii) inference-based watermarking \cite{lee-etal-2024-wrote} \cite{pmlr-v202-kirchenbauer23a} adapts the ``green/red lists'' idea from natural-language watermarking to code, which divides the token vocabulary (or a subset) into ``green'' and ``red'' lists for low-entropy tokens, and biases the decoder during inference to sample primarily from the green list. This reduces the chance of accidentally altering critical syntax while embedding the watermark, and (iii) neural-based watermarking~\cite{yang2023towards} takes a learning-based, end-to-end approach, which embeds watermark signatures into generated code via a neural insertion module and relies on structural transformations to preserve the abstract syntax tree (AST) invariants to maintain correct functionality. 


\subsection{Reinforcement Learning for LLM alignment}
Reinforcement learning (RL) has become a central component in post-training large language models beyond supervised finetuning \cite{ouyang2022traininglanguagemodelsfollow}. In particular, reinforcement learning from human feedback (RLHF) aligns model outputs with human preferences, improve reasoning, and optimize for task-specific objectives that are difficult to capture through static datasets. A typical RLHF pipeline builds upon a reward signal, which is derived from human feedback, learned reward models, or rule-based evaluators, and a policy model using variants of policy-gradient methods. Despite its effectiveness, RLHF pipeline can be computationally expensive, tightly coupled to reward-model quality, and prone to reward hacking or distributional drift.

Group Relative Policy Optimization (GRPO) offers a lightweight alternative by leveraging relative comparisons within a group of sampled responses. Given a prompt $x$ and a set of candidate outputs $\{y_{1},\dots,y_{K}\}$ sampled from the current policy $\pi_{\theta}$, each response is assigned a score $s_{k}$ from an evaluator (e.g., functional compiler for coding tasks). The relative advantage is computed as:
\[
A_{k} = s_{k} - \frac{1}{K}\sum_{j=1}^{K} s_{j},
\]
The policy update then maximizes the objective
\[
\mathcal{L}_{\text{GRPO}}(\theta) = \frac{1}{K}\sum_{k=1}^{K} A_{k} \log \pi_{\theta}(y_{k}\mid x).
\]

\section{Methodology}

\begin{figure}[!ht]
    \centering
    \begin{minipage}[c]{0.62\textwidth}
        \centering
        \includegraphics[width=\linewidth]{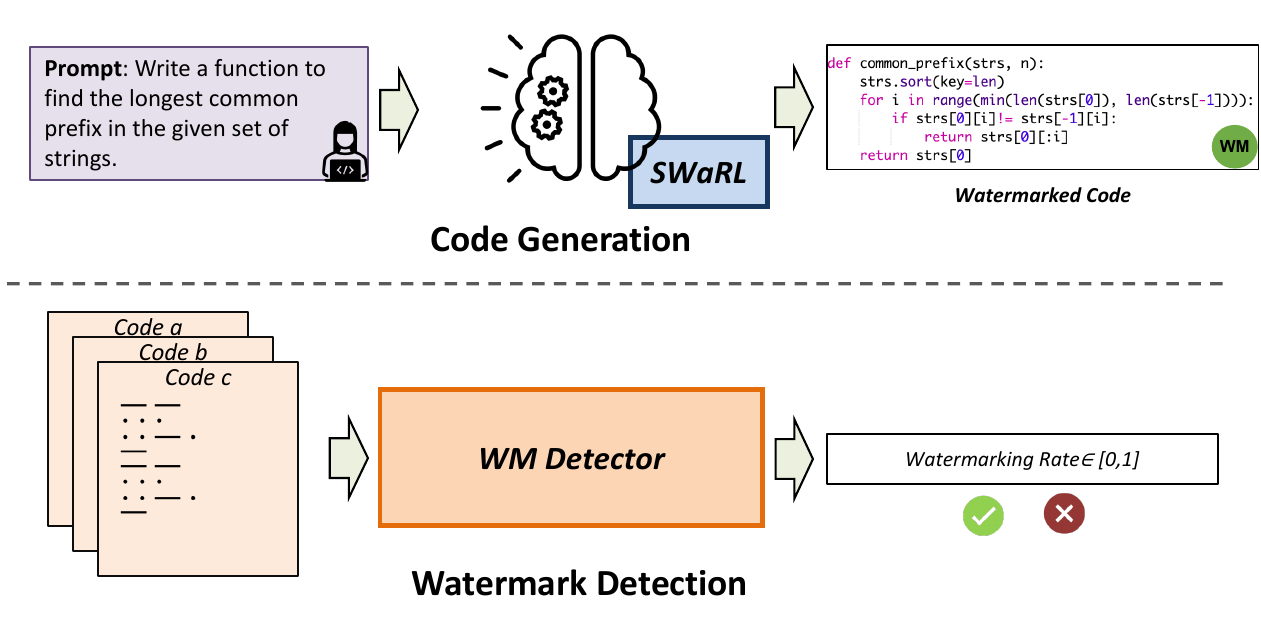}
    \end{minipage}
    \hfill
    \begin{minipage}[c]{0.34\textwidth}
        \caption{A code LLMs is fine-tuned with \sys{} to generate functional code with detectable watermark signatures. During deployment, users interact through a cloud API, which generates code containing watermark signatures. The model owner can then use the detector to verify whether a suspect code is watermarked or not.}
        \label{fig:fig1}
    \end{minipage}
\end{figure}

\paragraph{Threat model}As shown in Figure~\ref{fig:fig1}, we consider the code LLM owner hosts the model as a cloud API service. Users receive only the final watermarked code, and have no access to the code generation and watermarking process. An adversarial user may attempt to tamper with or transform the watermarked code through post-processing and refactoring to remove the embedded watermarking signature. However, any manipulated code must remain functional and semantically valid, as the adversary relies on it for normal use or distribution. 

The attacker is assumed to have full access to the watermarked code produced by the provider, including the ability to read, edit, refactor, and execute it. However, the attacker does not have access to the provider’s watermarking key or insertion and verification mechanisms.
The attacker’s primary objective is to deactivate the watermark verification process by preventing correct message recovery, while preserving the functional behavior of the code. The attacker therefore aims to strike a balance between aggressive code manipulation and maintaining correctness, efficiency, and compatibility requirements.






\subsection{\sys{} Overview}
\begin{figure*}[h!]
    \centering
    \includegraphics[width=\linewidth]{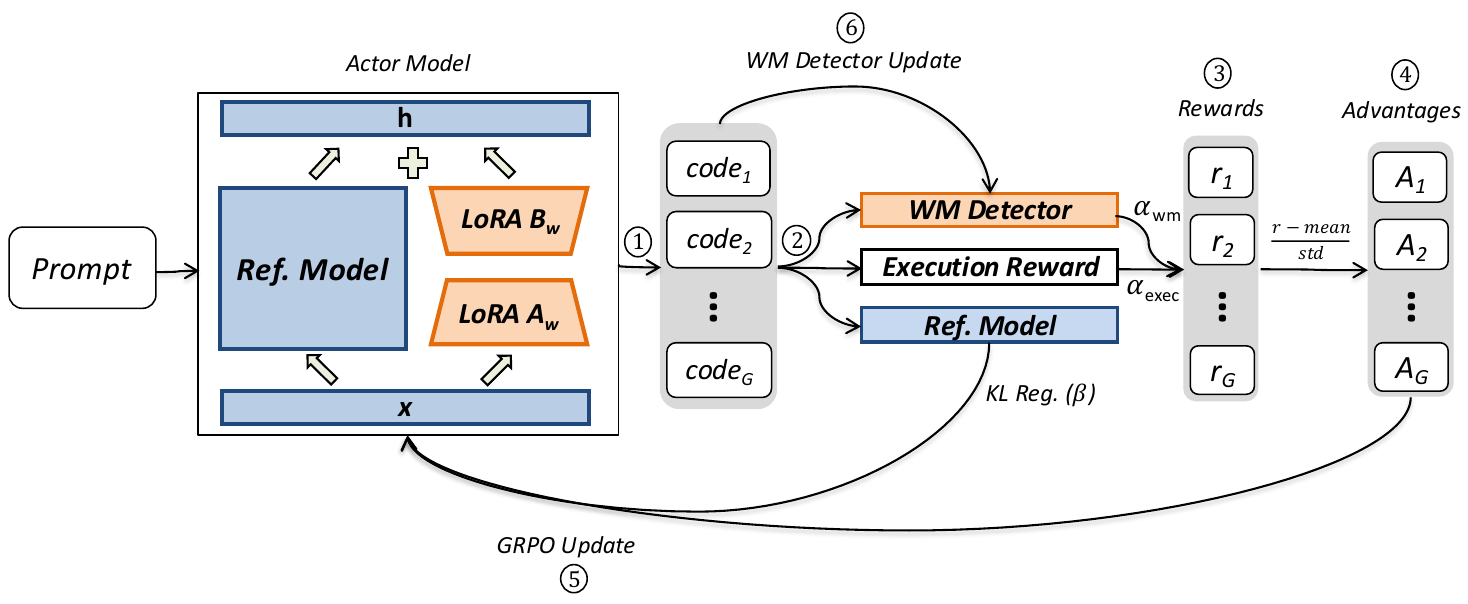}
    \caption{\sys{} training pipeline. The actor model generates candidate code, which is then evaluated both by a watermark detector (encouraging embedded watermark signals) and by execution tests (ensuring functional correctness). The combined reward drives a Group Relative Policy Optimization (GRPO) update, adjusting the model via lightweight Low-Rank Adaptation (LoRA). Periodically, the detector is retrained on new outputs so it stays aligned with the evolving actor policy.} 
    \label{fig:fig2}
\end{figure*}

\sys{} consists of two primary components: an actor model for generating candidate code, and a reward function that validates if the generated code is functional and have watermark embedded. These two components are aligned end-to-end as in shown Figure~\ref{fig:fig2}.

During fine-tuning, \sys{} uses GRPO-based RL optimization with a hybrid objective that jointly encourages watermark detectability and execution correctness, while constraining divergence from the reference model for stable training. The watermark detector is periodically co-trained with the actor to remain aligned with the evolving generation policy.

At deployment time, the fine-tuned LLM can be deployed through standard cloud APIs and behaves identically to conventional code-generation services, producing natural and executable code. Authorized parties may then apply the watermark detector to verify whether any given code snippet originated from the watermarked actor, enabling reliable provenance tracking without altering user workflow or output quality.

\subsection{GRPO-based optimization}
To train the actor code LLM to produce code that is both watermarked and functionally correct, we extend GRPO to optimize a hybrid reward combining three objectives: (1) a watermark detection score that encourages embedding model‑specific watermark signals, (2) an execution‑based correctness score that ensures watermarking does not compromise functional correctness and (3) a divergence penalty that constrains the policy to stay close to the original reference model. Concretely, for each sampled trajectory (code output) \(\tau\), we define  the scores as in Equation~\ref{eq:score}, where \(R_{\text{wm}}\) is the watermark reward, \(R_{\text{exec}}\) is the test‑execution reward, and the KL term penalizes divergence from the reference policy \(\pi_{\text{ref}}\).  

\begin{equation}
\begin{split}
   R(\tau) = \lambda_{\text{wm}}\,R_{\text{wm}}(\tau) \;+\; \lambda_{\text{exec}}\,R_{\text{exec}}(\tau)
    \;-\
    \beta\,D_{\mathrm{KL}}\bigl(\pi_\theta(\cdot\mid x)\,\big\|\,\pi_{\text{ref}}(\cdot\mid x)\bigr),
   \label{eq:score}
\end{split}
\end{equation} 

Given a group of \(G\) trajectories \(\{\tau_i\}_{i=1}^G\) with rewards \(R_i = R(\tau_i)\), GRPO computes the group mean \(\mu\) and standard deviation \(\sigma\), and defines the normalized advantage as \(A_i = \frac{R_i - \mu}{\sigma}\).
The actor is then updated via a clipped policy-gradient objective as:  
\begin{equation}
\mathcal{L}_{\text{clip}}(\theta)
= -\frac{1}{G} \sum_{i=1}^{G} \sum_{t=1}^{T} 
\min\!\Bigl(r_{i,t} A_i,\;\mathrm{clip}(r_{i,t}, 1-\epsilon, 1+\epsilon)\,A_i\Bigr)
\label{eq:grpo_clip}
\end{equation}
where \(r_{i,t} = \dfrac{\pi_\theta(a_{i,t} \mid a_{i,<t}, x)}{\pi_{\theta_{\text{old}}}(a_{i,t} \mid a_{i,<t}, x)}\) is the importance ratio, and \(\epsilon\) is the trust‑region (clipping) parameter.

As such, the actor learns to prefer outputs that balance watermark embedding with functional correctness, while the KL regularization term prevents the policy from drifting too far from the original model, which lead to stable and controlled learning of watermarked code generation.  

To make the GRPO fine-tuning procedure computationally efficient and the watermarked behavior modular, \sys{} adopts Low-Rank Adaptation (LoRA)~\cite{hu2022lora}. Rather than updating all parameters of the actor model, we inject small trainable low-rank matrices into its linear layers (the attention and MLP layers). Concretely, for any weight matrix \(W_0 \in \mathbb{R}^{d_{\text{out}} \times d_{\text{in}}}\), we reparameterize the updated weight as \(W = W_0 + BA\), where \(A \in \mathbb{R}^{r \times d_{\text{in}}}\) and \(B \in \mathbb{R}^{d_{\text{out}} \times r}\), and rank \(r \ll \min(d_{\text{in}}, d_{\text{out}})\). Only the adapter matrices \(A,B\) are trainable, while the original base weights \(W_0\) remain frozen throughout training.
\subsection{Reward objective}
During GRPO fine‑tuning, each generated code sample (trajectory) \(\tau\) is scored with a hybrid reward that balances three objectives: embedding a detectable watermark as $R_{\mathrm{wm}}$,  preserving functional correctness as $R_{\mathrm{exec}}(\tau)$, and a KL divergence penality. 

\paragraph{Watermark-based reward.}  
The watermark-score \(R_{\mathrm{wm}}(\tau)\) is obtained from a frozen detector model \(D(\cdot; \theta_d)\), which estimates how likely \(\tau\) was generated by the actor model. Thus, we formulate the score in Equation~\ref{eq:detect_score}.

\begin{equation}
    R_{\mathrm{wm}}(\tau) = D(\tau; \theta_d).
\label{eq:detect_score}
\end{equation}

To keep \(D\) aligned with the evolving actor policy, we periodically unfreeze and fine‑tune the detector: given a batch \(\mathcal{B}_{\mathrm{det}}\) of recent actor‑generated samples (labeled as positive source) and external/reference samples (labeled as non‑source), we minimize the binary cross‑entropy loss  as in Equation~\ref{eq:loss}, where \(y_{\mathrm{src}} = 1\) for actor outputs and \(0\) for reference/external code.

\begin{equation}
\begin{split}
  \mathcal{L}_{\mathrm{det}}(\theta_d) = -\frac{1}{|\mathcal{B}_{\mathrm{det}}|} \sum_{\tau \in \mathcal{B}_{\mathrm{det}}} \bigl[y_{\mathrm{src}} \log D(\tau;\theta_d) + (1 - y_{\mathrm{src}}) \log (1 - D(\tau;\theta_d))\bigr]
   \label{eq:loss}
\end{split}
\end{equation}

\paragraph{Execution-based reward.}  
The execution score \(R_{\mathrm{exec}}(\tau)\) measures functional correctness by compiling and running the generated code against a suite of \(K\) unit tests as in Equation~\ref{eq:exe_score}.
\begin{equation}
    R_{\mathrm{exec}}(\tau) = \frac{1}{K} \sum_{k=1}^{K} \mathbb{1}[\mathrm{test}_k(\tau) = \mathrm{pass}]
    \label{eq:exe_score}
\end{equation}

\medskip  
By combining \(R_{\mathrm{wm}}\) and \(R_{\mathrm{exec}}\) in the hybrid reward, GRPO training encourages the actor to generate code that is both watermarked (detectable by \(D\)) and functionally correct.

\subsection{End-to-end \sys{} fine-tuning}

\noindent
\begin{minipage}[t]{0.48\linewidth}
\vspace{0pt}
We now show the full training loop of \sys{} in Algorithm~\ref{alg:sys-finetune}, which alternates between policy optimization of the code LLM (actor) and periodic updates to the watermark detector. The loop is initialized with actor parameters $\theta_s$ from an SFT checkpoint and detector parameters $\theta_0$ pretrained to distinguish SFT outputs from reference data. In each iteration, a mini-batch of prompts is sampled; for each prompt, multiple code candidates are generated. Each candidate receives a hybrid reward combining a {\em watermark score} (from the current detector) and an {\em execution correctness score}. The actor is updated using a group-relative policy gradient, where the group baseline normalizes the rewards across all candidates. Every \(K\) steps, the detector is co-optimized on a mix of actor-generated and reference code, to maintain its ability to discriminate watermarked outputs from others. Over time this procedure yields a code LLM that consistently produces functional, watermarked code, while keeping the detector aligned with the evolving policy.
\end{minipage}
\hfill
\begin{minipage}[t]{0.50\linewidth}
\vspace{-3pt}
\begin{algorithm}[H]
\caption{Full \sys{} training loop}
\label{alg:sys-finetune}
\DontPrintSemicolon
\KwIn{Actor $\theta_w \leftarrow \theta_s$, detector $\theta_d \leftarrow \theta_0$, dataset $\mathcal{D}$, group size $G$, steps $S$, detector interval $K$, reference policy $\pi_{\text{ref}}$, learning rates $\eta_w,\eta_d$}
\KwOut{Trained actor $\theta_w^*$ and detector $\theta_d^*$}
\For{$s = 1$ \KwTo $S$}{
    Sample $\mathcal{B} \subset \mathcal{D}$\;

    \ForEach{$(x,y_{\text{ref}},tc) \in \mathcal{B}$}{ \tcp{Actor update}
        Sample $\{\tau^{(g)}\}_{g=1}^{G} \sim \pi_{\theta_w}(\cdot \mid x)$\;
        Compute $R(\tau^{(g)})$ and $\mathcal{L}_{\text{clip}}$ (Eq.~\ref{eq:score},~\ref{eq:grpo_clip})\;
    }
    $\theta_w \leftarrow \theta_w - \eta_w \nabla_{\theta_w}
    \mathbb{E}_{(x,\cdot,\cdot)\sim\mathcal{B}} \mathcal{L}_{\text{clip}}(x)$\;

    \If{$s \bmod K = 0$}{ \tcp{Detector update}
        Sample $\mathcal{B}_d$\;
        Compute $\mathcal{L}_{\mathrm{det}}$ (Eq.~\ref{eq:loss})\;
        $\theta_d \leftarrow \theta_d - \eta_d \nabla_{\theta_d}
        \mathcal{L}_{\mathrm{det}}(\theta_d;\mathcal{B}_d)$\;
    }
}
\Return $\theta_w^*, \theta_d^*$\;
\end{algorithm}
\end{minipage}

\subsection{\sys{} deployment}
During inference, the fine-tuned LLM is deployed as a standard code-generation API or service. The SFT-trained base model and the GRPO-tuned LoRA adapter are merged so that the model produces watermark-embedded code. Each time the API receives a prompt, the fine-tuned LLM generates code that (i) is functional to pass unit tests, and (ii) carries the signature that \sys{} has learned.  The trained watermark detector can be later used to verify if a code is watermarked. Given any arbitrary code sample, the watermark detector outputs a likelihood indicating whether the code was generated by code LLM trained with \sys{}. This enables external auditing or provenance checking without needing access to the original LLM or retraining. 

\section{Experiments}
\subsection{Experimental setup}
\paragraph{Training details} \sys{} follows a three-stage pipeline consisting of SFT, detector pretraining, and GRPO-based co-training, where each stage uses a separate split of the \texttt{OpenCoder-LLM} ~\cite{Huang2024OpenCoderTO} dataset. We initialize the actor from \texttt{Qwen/Qwen2.5-Coder-1.5B-Instruct} ~\cite{qwen2025qwen25technicalreport} and apply LoRA-based SFT. The detector is implemented using \texttt{microsoft/codebert-base} \cite{feng2020codebert} with maximum input length $512$, and is pretrained on $10{,}000$ SFT generated samples paired with reference code. During GRPO training, a new LoRA adapter is attached to the SFT actor and optimized on $10{,}000$ samples using up to four completions per prompt, temperature $0.9$, top-$p=0.95$, maximum generation length $2048$, and KL coefficient $0.05$. The final reward combines watermark and execution rewards weighted by $\alpha_{\text{wm}}=0.4$ and $\alpha_{\text{exec}}=0.6$, respectively. Training is performed on a node with eight AMD Instinct MI250 GPUs.

\paragraph{Evaluation setup} All experiments are conducted on the \texttt{mi210x1} GPU partition equipped with AMD Instinct MI210 GPUs. For fair comparison, all methods use the same SFT-trained \texttt{Qwen/Qwen2.5-Coder-1.5B-Instruct} backbone. For each benchmark prompt, responses are generated with maximum sequence length $2048$, temperature $0.2$, top-$p=0.95$, and batch size $40$. We evaluate five systems: the SFT baseline, EXP-edit~\cite{kuditipudi2023robust}, SWEET~\cite{lee-etal-2024-wrote} ($\delta=3$, $\gamma=0.25$), WLLM~\cite{li2025acw} ($\delta=3$, $\gamma=0.25$), and the proposed \sys{}.

\paragraph{Metrics and datasets} We evaluate both functional correctness and watermark detectability on human-written and LLM-generated code. Functional performance is measured using \textit{Pass@k}, which estimates the probability that at least one generated solution passes the test cases within $k$ attempts, while watermark detectability is measured using \textit{AUROC} (Area Under the ROC Curve), a threshold-independent metric for distinguishing watermarked from non-watermarked samples. Experiments are conducted on HumanEval~\cite{chen2021evaluating} and MBPP~\cite{austin2021program}, along with their extended variants HumanEval+ and MBPP+~\cite{liu2023your}.

\subsection{Main results}~\label{subsec:main_results}
\begin{figure}[t]
\vspace{-10pt}
    \centering
    \begin{minipage}[t]{0.58\linewidth}
        \centering
        \includegraphics[width=\linewidth]{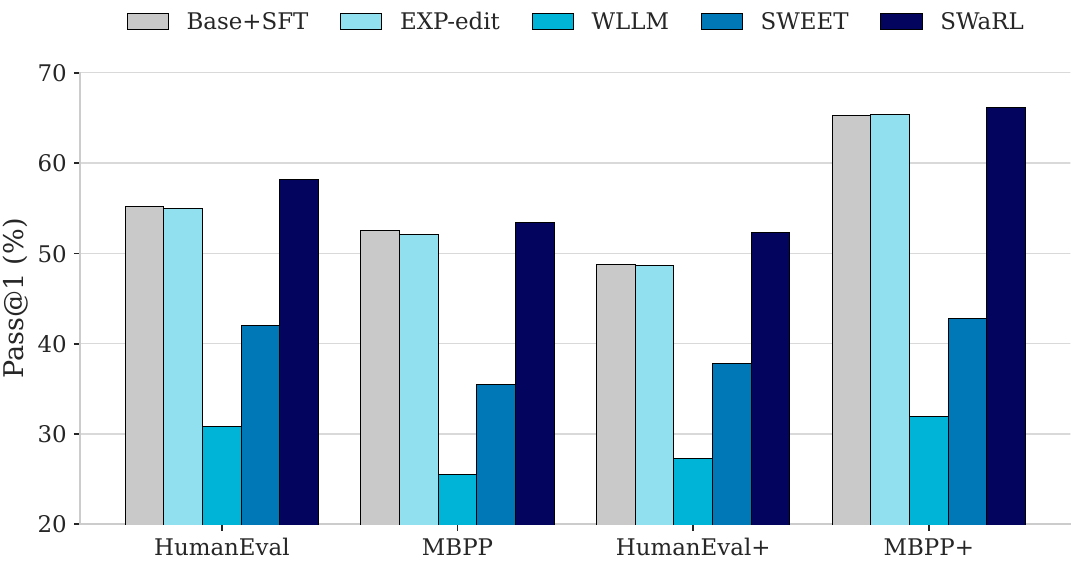}
        \vspace{-4pt}
        \subcaption{}
        \label{fig:pass1_barplots}
    \end{minipage}
    \hspace{-5pt}
    \begin{minipage}[t]{0.4\linewidth}
        \centering
        \includegraphics[width=\linewidth]{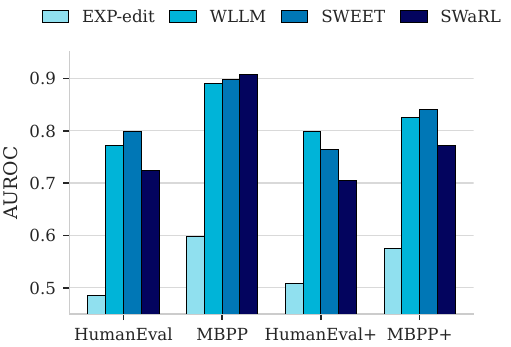}
        \vspace{-4pt}
        \subcaption{}
        \label{fig:auroc_barplots}
    \end{minipage}
    
    \vspace{-2pt}
    \caption{Comparison of watermarking methods across HumanEval, MBPP, HumanEval+, and MBPP+ datasets. (a) Pass@1 comparison showing that \sys{} consistently achieves higher functional correctness across benchmarks. (b) AUROC comparison showing that \sys{} maintains strong watermark detectability while preserving code quality.}
    \label{fig:main_results_barplots}
    \vspace{-10pt}
\end{figure}

Table~\ref{tab:watermark_results} presents the functional correctness and watermark detectability results across HumanEval, MBPP, HumanEval+, and MBPP+. Functional performance is evaluated using Pass@1 and Pass@10, while watermark detectability is measured using AUROC. Figure~\ref{fig:main_results_barplots} further visualizes the tradeoff between code quality and watermark detectability across methods.

\begin{table*}[!ht]
\centering
\caption{Main results of code-generation accuracy and watermark-detection performance across four benchmarks. Best results among watermarking methods are shown in \textbf{bold}; second-best are \underline{underlined}.}
\scriptsize
\setlength{\tabcolsep}{3pt}
\renewcommand{\arraystretch}{0.95}
\begin{tabular}{l ccc ccc ccc ccc}
\toprule
\textbf{Method} 
& \multicolumn{3}{c}{\textbf{HumanEval}} 
& \multicolumn{3}{c}{\textbf{MBPP}}
& \multicolumn{3}{c}{\textbf{HumanEval+}}
& \multicolumn{3}{c}{\textbf{MBPP+}}\\
& Pass@1 & Pass@10 & AUROC 
& Pass@1 & Pass@10 & AUROC
& Pass@1 & Pass@10 & AUROC
& Pass@1 & Pass@10 & AUROC\\
\midrule
\multicolumn{13}{l}{\textbf{No Watermark}} \\
Base+SFT        
            & 55.23 & 69.28 & - 
            & 52.52 & 71.95 & -
            & 48.81 & 62.46 & -
            & 65.26 & 84.18 & -\\

\midrule
\multicolumn{13}{l}{\textbf{Watermarks}} \\

EXP-edit    
            & \underline{55.00} & \underline{69.09} & 0.486 
            & 52.05 & \underline{70.97} & 0.598
            & 48.63 & \underline{62.60} & 0.508
            & \underline{65.38} & \underline{82.90} & 0.575 \\

WLLM        
            & 30.82 & 51.07 & \underline{0.771}
            & 25.54 & 52.16 & 0.890
            & 27.24 & 45.20 & \textbf{0.799}
            & 31.93 & 63.31 & \underline{0.825} \\

SWEET       
            & 42.01 & 58.49 & \textbf{0.798}
            & 35.49 & 56.00 & \underline{0.898}
            & 37.82 & 54.32 & \underline{0.765}
            & 42.79 & 63.29 & \textbf{0.841} \\

\sys{} 
            & \textbf{58.22} & \textbf{78.64} & 0.725
            & \textbf{53.40} & \textbf{70.97} & \textbf{0.908}
            & \textbf{52.36} & \textbf{70.53} & 0.705
            & \textbf{66.20} & \textbf{83.03} & 0.772 \\

\bottomrule
\end{tabular}
\label{tab:watermark_results}
\end{table*}

The results show that prior watermarking methods struggle to simultaneously maintain strong detectability and high code-generation quality, whereas \textbf{\sys{} achieves a better balance of detectability and fidelity-preservation}. Statistical token-level watermarking approaches such as WLLM and SWEET obtain strong AUROC scores across most benchmarks, but significantly degrade functional correctness relative to the SFT baseline. For example, on HumanEval, WLLM improves AUROC to $0.771$ while reducing Pass@1 from $55.23$ to $30.82$. SWEET exhibits a similar trend, achieving AUROC values above $0.79$ on multiple benchmarks while consistently lowering Pass@1 and Pass@10. In contrast, EXP-edit better preserves generation quality, but provides substantially weaker watermark detectability, with AUROC values near random guessing on HumanEval and HumanEval+.

Compared to prior methods, \sys{} achieves the strongest overall balance between functionality preservation and watermark detection. As shown in Figure~\ref{fig:main_results_barplots}(b), \sys{} consistently attains the highest Pass@1 among all watermarking methods across all four benchmarks. Meanwhile, Figure~Figure~\ref{fig:main_results_barplots}(a) shows that \sys{} maintains strong watermark detectability, including the best AUROC on MBPP ($0.908$). Unlike token-level statistical watermarking methods that directly perturb the decoding distribution, \sys{} learns watermark-consistent generation behaviors through RL-based optimization, enabling robust watermark embedding without severely compromising functional correctness.

Another notable observation is that \sys{} not only preserves utility under watermarking, but also surpasses the non-watermarked SFT baseline in several settings. For example, on HumanEval, \sys{} improves Pass@1 from $55.23$ to $58.22$ and Pass@10 from $69.28$ to $78.64$. Similar improvements are observed on HumanEval+ and MBPP+, where \sys{} consistently exceeds the Base+SFT model in functional correctness. These gains suggest that the GRPO objective, together with execution-based rewards, acts as an additional functionality-alignment stage that reinforces syntactically valid and semantically correct program generation beyond supervised fine-tuning alone.

    

    

\subsection{Robustness}~\label{sec:robustness}
\subsubsection{Refactoring attack}

For the refactoring attack evaluation, we employ two third-party LLMs, Qwen/Qwen2.5-3B-Instruct \cite{qwen2025qwen25technicalreport} (\texttt{QW Attack}) and deepseek-ai/deepseek-coder-7b-instruct-v1.5 \cite{guo2024deepseekcoderlargelanguagemodel} (\texttt{DS Attack}) which are not involved in any stage of our training pipeline. The attacker model is prompted to refactor each generated code sample into an alternative, functionally equivalent implementation. These refactored outputs are then passed to the \sys{} watermark detector to assess watermark removability. This setup enables us to quantify how effectively an external model can erase or obfuscate the embedded watermark while preserving program semantics.

Table~\ref{tab:refactor_attack_auroc} reports AUROC before and after refactoring attacks on HumanEval and MBPP datasets. Existing watermarking methods exhibit substantial degradation under rewriting. WLLM experiences AUROC drops of $14$--$17\%$ across attacks and datasets, while SWEET shows similar instability, with degradation exceeding $11\%$ in all settings and reaching $16.82\%$ on MBPP under \texttt{DS Attack}. Although EXP-edit shows smaller degradation, its watermark detectability is already weak before attack, limiting its practical robustness.

In contrast, \sys{} consistently demonstrates stronger robustness under adversarial rewriting. Under \texttt{QW Attack}, \sys{} achieves AUROC reductions of only $3.45\%$ on HumanEval and $9.40\%$ on MBPP. Under the stronger \texttt{DS Attack}, degradation further decreases to $5.66\%$ and $2.21\%$, respectively. Moreover, \sys{} retains the highest post-attack AUROC on MBPP under both attacks ($0.819$ and $0.884$), demonstrating stronger resilience to semantics-preserving refactoring than prior token-level watermarking methods. 

\begin{table}[t]
\centering
\caption{
AUROC before and after adversarial refactoring attacks using Qwen2.5-3B-Instruct (\texttt{QW Attack}) and DeepSeek-Coder-7B-Instruct (\texttt{DS Attack}). $\Delta\downarrow$ denotes relative AUROC degradation (\%). Higher AUROC and lower degradation indicate stronger robustness.
}
\vspace{3pt}
\small
\setlength{\tabcolsep}{3pt}
\begin{tabular}{lcccccc}
\toprule
\textbf{Method} 
& \multicolumn{2}{c}{\textbf{No Attack}} 
& \multicolumn{2}{c}{\textbf{QW Attack}} 
& \multicolumn{2}{c}{\textbf{DS Attack}} \\
\cmidrule(lr){2-3} \cmidrule(lr){4-5} \cmidrule(lr){6-7}
& HumanEval & MBPP 
& HumanEval / $\Delta\downarrow$ & MBPP / $\Delta\downarrow$
& HumanEval / $\Delta\downarrow$ & MBPP / $\Delta\downarrow$ \\
\midrule

EXP-edit   
& 0.486 & 0.598 
& 0.475 / \textbf{2.26} & 0.556 / 7.02 
& 0.460 / \textbf{5.35} & 0.563 / 5.85 \\

WLLM       
& 0.771 & 0.890 
& 0.662 / 14.14 & 0.738 / 17.08 
& 0.663 / 14.01 & 0.765 / 14.04 \\

SWEET      
& \textbf{0.798} & 0.898 
& \textbf{0.704} / 11.78 & 0.787 / 12.36 
& \textbf{0.707} / 11.41 & 0.747 / 16.82 \\

\textbf{SWaRL} 
& 0.725 & \textbf{0.904} 
& 0.700 / 3.45 & \textbf{0.819} / \textbf{9.40} 
& 0.684 / 5.66 & \textbf{0.884} / \textbf{2.21} \\

\bottomrule
\end{tabular}
\label{tab:refactor_attack_auroc}
\end{table}

\subsubsection{Variable renaming attack}
\begin{figure}[t]
    \centering
    \begin{minipage}[c]{0.48\linewidth}
        \centering
        \includegraphics[width=\linewidth]{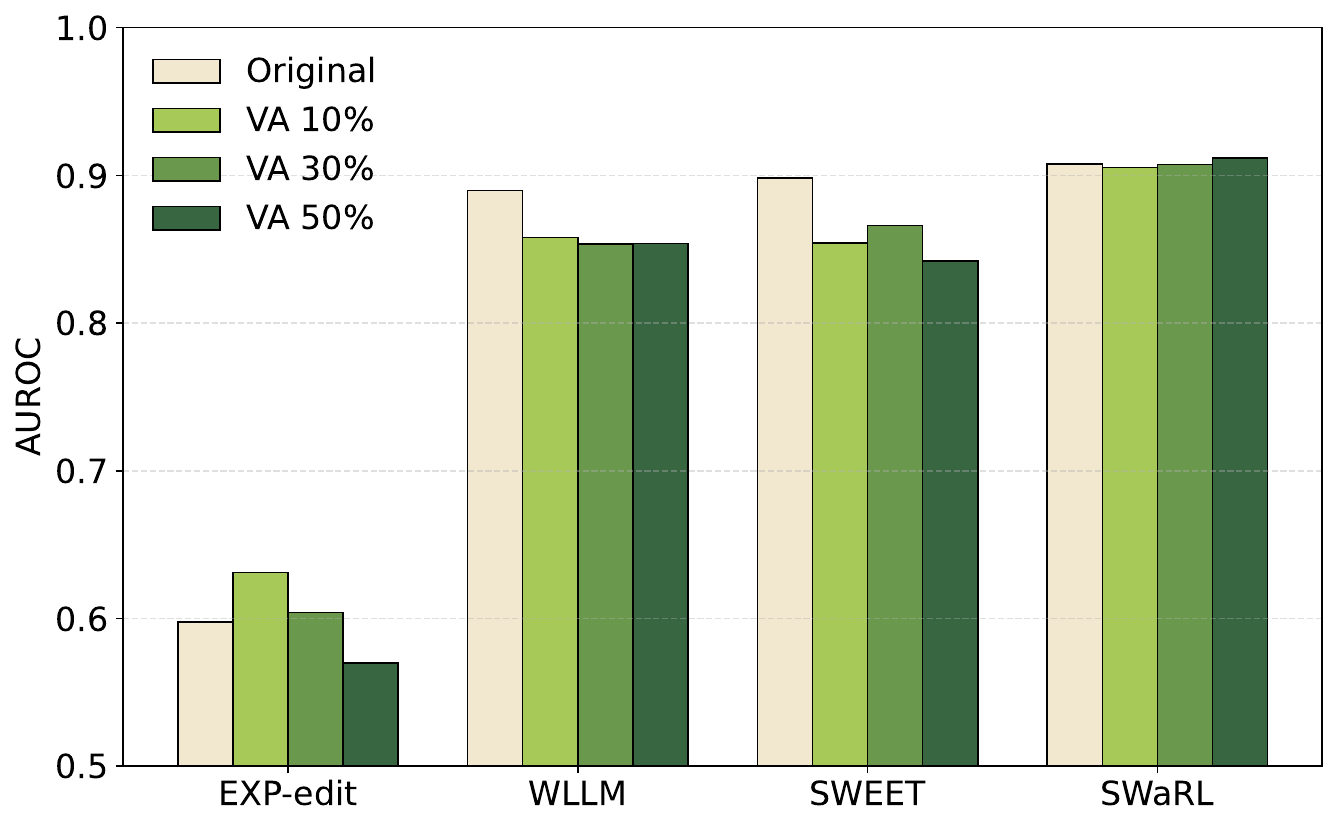}
    \end{minipage}
    \hspace{4pt}
    \begin{minipage}[c]{0.40\linewidth}
        \vspace{0pt}
        \caption{AUROC under variable renaming attacks (\texttt{VA}) on MBPP with 10\%, 30\%, and 50\% identifier perturbation ratios. \sys{} maintains consistently high detectability under increasing syntactic perturbations.}
        \label{fig:mbpp_renaming_auroc}
    \end{minipage}
    \vspace{-8pt}
\end{figure}
We further evaluate robustness against syntactic transformations using a variable renaming attack on the MBPP dataset. For each generated program, we randomly replace 10\%, 30\%, or 50\% of identifiers with variable names drawn from a pool of human-written code, while preserving program functionality and syntactic correctness.

Figure~\ref{fig:mbpp_renaming_auroc} shows that prior watermarking methods are sensitive to identifier-level transformations. WLLM and SWEET exhibit clear AUROC degradation as the renaming ratio increases, indicating reliance on lexical or token-level patterns. In contrast, \sys{} maintains consistently high AUROC across all perturbation levels, demonstrating stronger robustness to surface-level syntactic changes.

\subsection{Latency}\label{sec:latency}
\begin{wraptable}{r}{0.5\textwidth}
\vspace{-15pt}
\centering
\caption{Per-token generation and watermark-detection latency (seconds/token), where lower is better.}
\label{tab:latency_table}
\scriptsize
\setlength{\tabcolsep}{3.5pt}
\begin{tabular}{lcccc}
\toprule
& \textbf{EXP-edit} & \textbf{WLLM} & \textbf{SWEET} & \textbf{\sys{}} \\
\midrule
\textbf{Generation} $\downarrow$ 
& 0.023 & 0.040 & 0.044 & 0.039 \\
\textbf{Detection} $\downarrow$
& 0.588 & 0.001 & 0.004 & 0.0002 \\
\bottomrule
\end{tabular}
\vspace{-15pt}
\end{wraptable}

Table~\ref{tab:latency_table} reports per-token generation and watermark-detection latency across watermarking methods. Overall, \sys{} introduces minimal inference overhead while achieving the fastest detection runtime among all baselines.

\paragraph{Generation latency.} \sys{} achieves generation latency comparable to statistical watermarking approaches, with $0.039$ seconds/token versus $0.040$ and $0.044$ for WLLM and SWEET, respectively. This suggests that the LoRA-based RL alignment preserves efficient decoding behavior without introducing substantial inference overhead.

\paragraph{Detection latency.} \sys{} achieves the lowest detection latency at only $0.0002$ seconds/token, substantially outperforming all baselines. In contrast, SWEET and WLLM incur higher runtime due to token-level statistical computations, while EXP-edit exhibits particularly high verification cost ($0.588$ seconds/token), limiting its practicality for large-scale or real-time deployment.

\section{Limitations and Future Work} \label{limitations}
\sys{} provides sufficient watermark strength while preserving functionality, robustness, and efficiency. To support real-world deployment, several extensions could further enhance its capabilities. First, extending \sys{} to watermark compiled (binary-level) code would improve detectability when source code is unavailable. Second, while \sys{} currently evaluates performance at the function level, repository-level watermarking remains unexplored. This setting introduces a significantly larger adversarial modification space with more structurally diverse changes, making repository-level code watermarking an important direction for future work.

\section{Conclusion}
This paper presents \sys{}, a reinforcement learning-based watermarking framework for code LLMs that jointly optimizes functional correctness and watermark detectability. By combining GRPO-based alignment with lightweight LoRA adaptation, the framework enables efficient integration of watermarking behavior into pretrained code models. Extensive evaluations demonstrate that \sys{} achieves a strong tradeoff between code correctness and watermark detectability, consistently outperforming prior watermarking approaches in functional performance while maintaining robust watermark verification. Furthermore, \sys{} exhibits strong resilience against adversarial transformations, including refactoring and variable renaming attacks, while introducing minimal inference and detection overhead. These results demonstrate the effectiveness of \sys{} as a practical and scalable solution for attribution and ownership verification in code generation systems.
\bibliographystyle{unsrt}
\bibliography{references}

\end{document}